\documentclass[final,3p,times,twocolumn]{elsarticle}
\usepackage{amssymb}

\usepackage{lineno}
\usepackage{xcolor}
\usepackage[english]{babel}
\usepackage{ulem}
\journal{arXiv}

\usepackage{amsmath}
\usepackage{amssymb}
\usepackage{mathrsfs}
\usepackage{mathtools}
\usepackage{tikz}

\makeatletter
\newcommand\mathcircled[1]{%
  \mathpalette\@mathcircled{#1}%
}
\newcommand\@mathcircled[2]{%
  \tikz[baseline=(math.base)] \node[draw,circle,inner sep=1pt] (math) {$\m@th#1#2$};%
}
\newcommand{\xdashrightarrow}[3][]{\ext@arrow 0359\rightarrowfill@@{#1}{#2}}

\def\rightarrowfill@@{\arrowfill@@\relax\relbar\rightarrow}
\def\leftarrowfill@@{\arrowfill@@\leftarrow\relbar\relax}
\def\leftrightarrowfill@@{\arrowfill@@\leftarrow\relbar\rightarrow}
\def\arrowfill@@#1#2#3#4{%
  $\m@th\thickmuskip0mu\medmuskip\thickmuskip\thinmuskip\thickmuskip
   \relax#4#1
   \xleaders\hbox{$#4#2$}\hfill
   #3$%
}
\makeatother

\begin{document}

\newcommand{\marco}[1]{{\color{blue} #1}}
\newcommand{\mk}[1]{{\color{teal} #1}}
\newcommand{\luca}[1]{{\color{magenta} #1}}
\newcommand{\workinprogress}[1]{{\color{orange} #1}}

\begin{frontmatter}
\title{Proximity: a recipe to break the outbreak}

\author{Marco Faggian}
\author{Michele Urbani\corref{cor1}}
\ead{mk.urbani.89@gmail.com}
\author{Luca Zanotto}

\begin{abstract}
\parindent 0 pt
We present a mobile app solution to help the containment of an epidemic outbreak by keeping track of possible infections in the incubation period. We consider the particular case of an infection which primarily spreads among people through proximal contact, via respiratory droplets. This smartphone application will be able to detect other devices in close proximity and list all the interactions in an anonymous and encrypted way. If an app user is tested positive and so is certified as infected, the application notifies immediately the potential contagion to the devices in the list and suggests to start a voluntary quarantine and undergo a medical test. We believe this solution may be particularly useful in the current COVID-19 pandemic and moreover could be used to prevent similar events in the future.
\par
\end{abstract}


\end{frontmatter}

\parindent 0 pt

\section{Introduction}

What we can learn from the recent COVID-19 pandemic is that we were not ready for this kind of eventuality.\cite{BillGates} Our society has reached a level of global interconnection which makes an epidemic outbreak an event very problematic to contain with traditional quarantine measures.
This means that a very contagious disease can easily spread around the globe putting health care systems under pressure, especially in countries where such systems are already in precarious conditions.\\
Therefore, it is necessary to find a way to effectively limit pandemics as soon as possible.\par
Nowadays, especially in the most technologically developed areas, a vast majority of people own a smartphone and have it constantly with themselves. This type of behavior is typical of our current lifestyle and can represent an advantage to contain the epidemic outbreak.\\
An infectious disease primarily spreads among people due to their proximity, via respiratory droplets coming from coughs and sneezes or via direct contact.\cite{WHO} \\
In order to slow down the outbreak, the best procedure is to immediately identify the "patient zero" and quarantine those who have come into contact with them. \cite{paz0}\\
The recent outbreak of COVID-19 showed that this approach is difficult to pursue and unsustainable when the incubation period is long, because of the high number of interactions and the lack of strategies to efficiently warn the people who could have gotten infected.
\par

The absence of an efficient method to track the spread of the epidemic may translate into an exponential growth of cases, and this, in turn, might lead to a pandemic which is hard to contain without severe restrictions on people's freedom.

\begin{figure}[htbp!]
\centering
\includegraphics[width=0.45\textwidth]{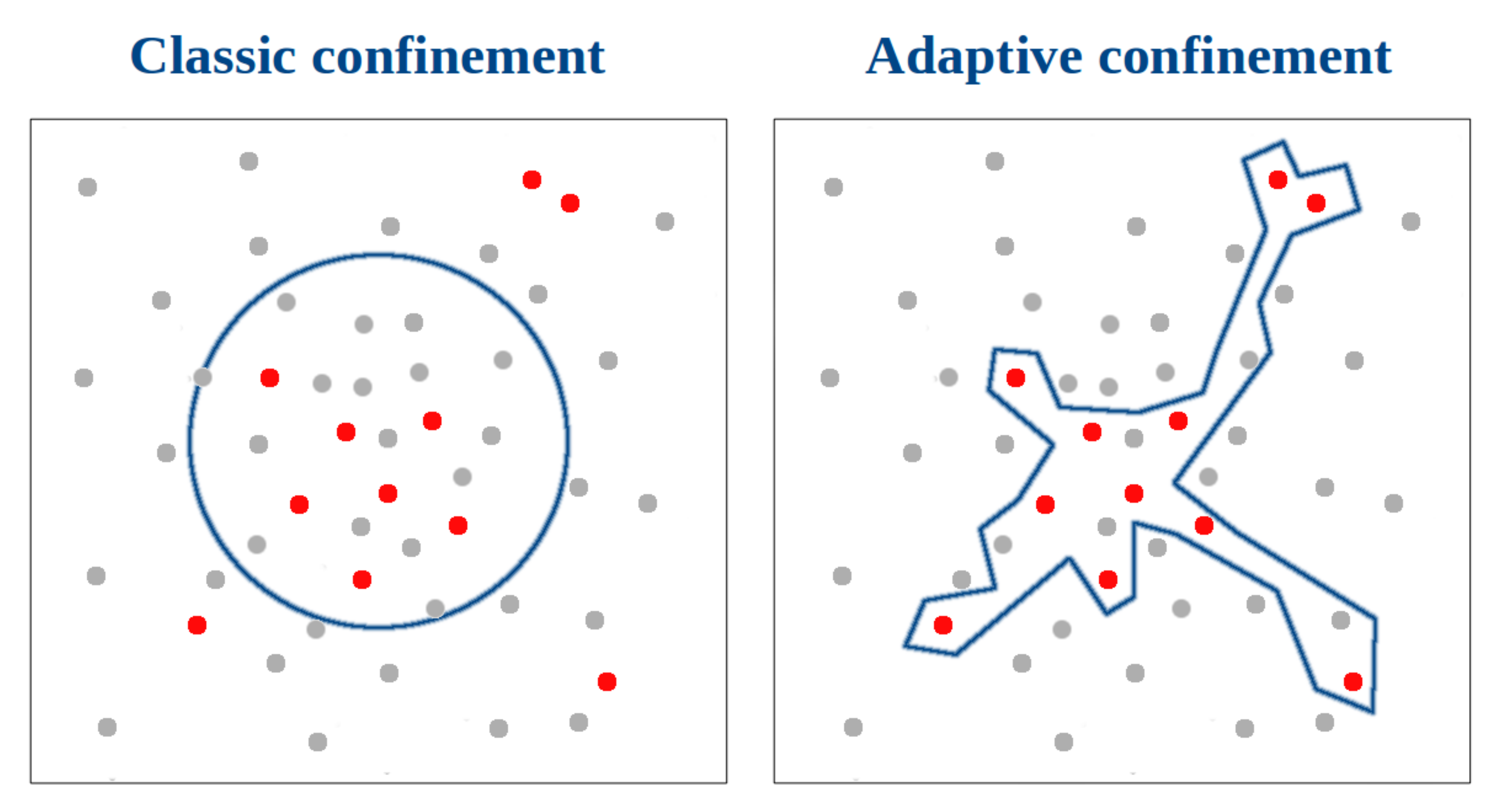}
\caption{Conceptual comparison between the adaptive confinement with respect to the classic confinement localized in a defined area.}
\label{confinement}
\end{figure}

We propose a relatively simple but powerful tool, based on an smartphone application, to statistically slow down and contain an epidemic outbreak.

The main idea is to treat each "positive patient" like the "patient zero" using a smartphone contact tracing strategy.
This approach can be called "selective quarantine" and contrasts with the "traditional quarantine" which consists in isolating a geographical area (see Fig\ref{confinement}).

This smartphone application should be able to track other devices in close proximity and to encrypt all the interactions into an anonymous "\textit{proximal contact list}". In the case where a patient is infected, the medical authority provides a key to activate the app, which sends automatic notifications to the devices on the list, warning about the possible contagion.

\section{Description of the Proximity app}

In this section we describe the workflow of the proposed smartphone application, that we named \textit{Proximity} \cite{Prox} , together with its main technical features and further development.

\subsection{ Main features}

The application works in two different modalities. One is the \textit{Tracking mode}, which runs in background and enables the recording of devices in close proximity. The other is the \textit{Alert mode}, which is activated only when the infection is identified and allows the device to send alerts to other potentially infected individuals.

\begin{figure}[htbp!]
\centering
\includegraphics[width=0.48\textwidth]{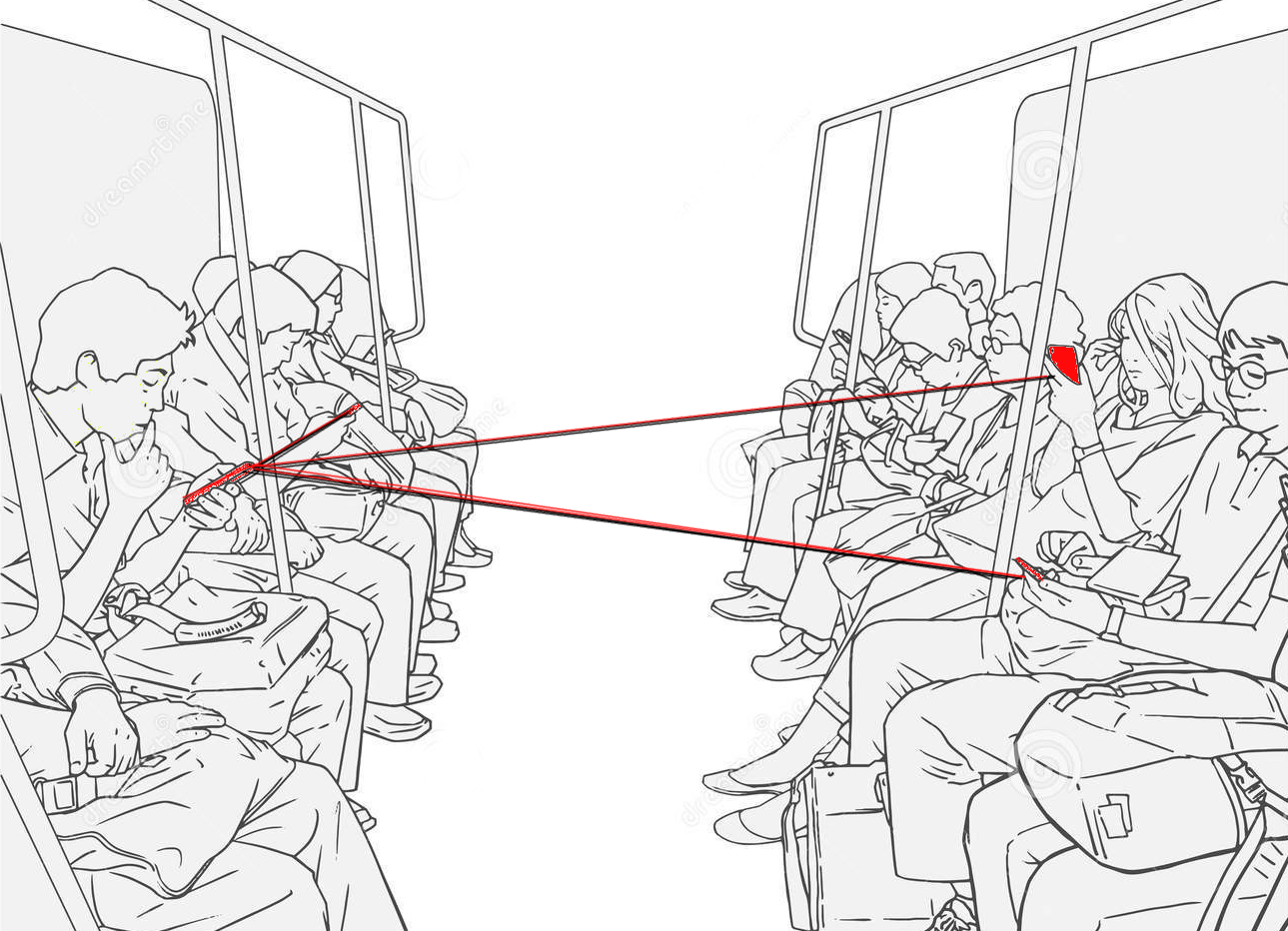}
\caption{Illustrative sketch which shows how Proximity should work and keep track of all short range (2-3 m) interactions.}
 \label{prox}
\end{figure}

\textit{Proximity}, in the \textbf{\textit{Tracking mode}}, should be able to:

\begin{enumerate}
 
 \item track all short range (distance of $ 2-3 m$) interactions between people with a smartphone with the same app installed, as illustrated in Fig.(\ref{prox}).
 
 \item save at least the following information: date, time and the duration of the interaction, and the phone-number of all the interacting mobile phones. 
 
 \item keep in the local memory a list (\textit{proximal contact list}) of all the interactions occurred in a time duration at least equal to the incubation period of the infection.
 
 \item be switched to the \textit{Alert mode}, by inserting an activation key.
 
\end{enumerate}

In the case a person is tested positive for the infection, the medical authority who certified the test, provides the key $K_m$ to activate the \textbf{\textit{Alert mode}}.
This key is a one-time password generated by an algorithm in a protected web page where only doctors with certified emails can access.
The medical authority can generate the key $K_m$ introducing the user ID, provided by the patient, in the online key generator.
Subsequently, the $K_m$ key is used by the patient to activate the \textit{Alert mode}.

\begin{figure*}[hbt!]
  \includegraphics[width=0.99\textwidth]{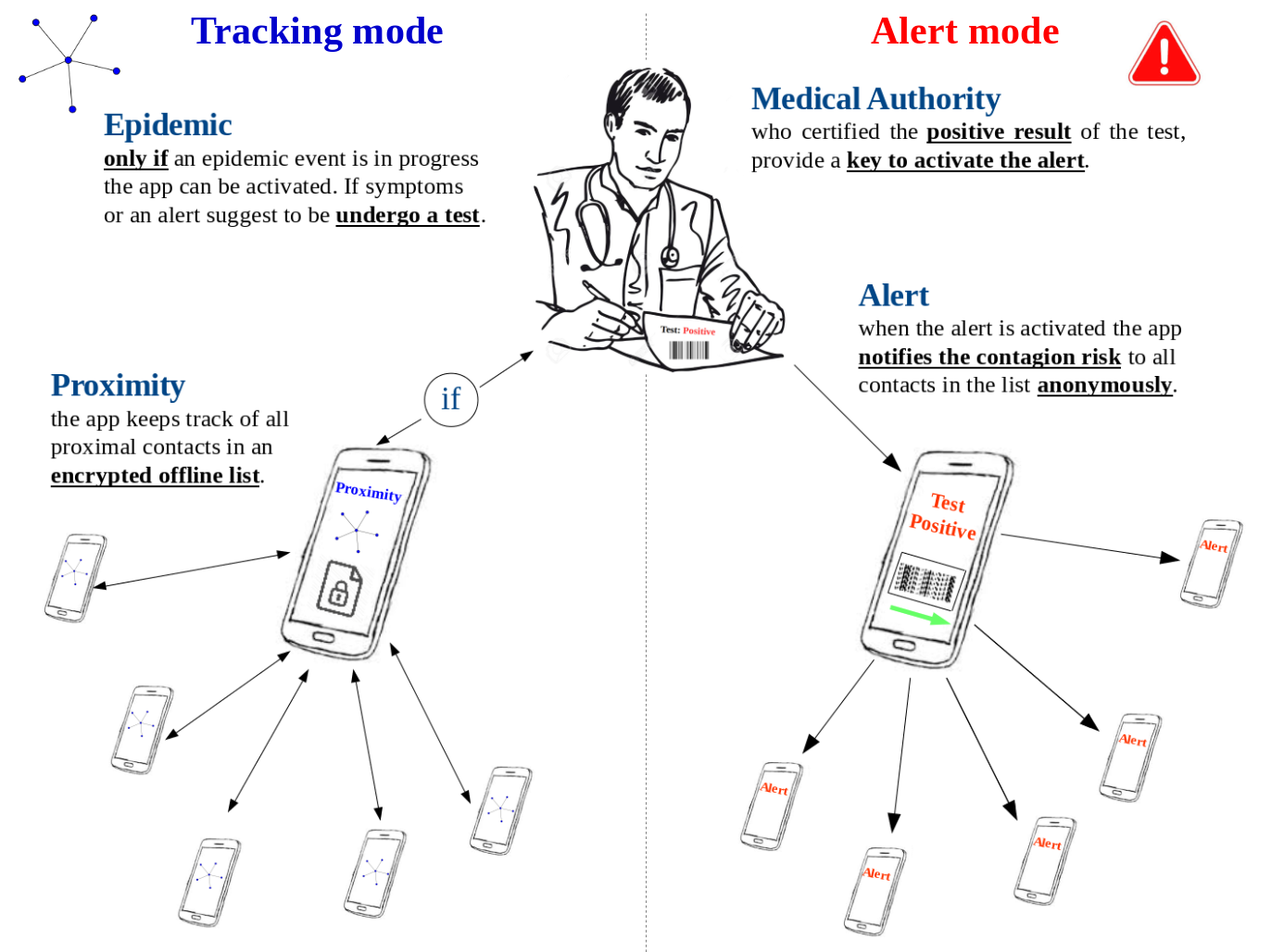}
    \caption{The diagram summarizes how \textit{Proximity} app works. On the left in the \textit{Tracking mode} and on the right in \textit{Alert mode}.}
    \label{diagram}
\end{figure*}

Once the Alert mode has been activated, the app will:
\begin{enumerate}

 \item send an SMS notification to all the contacts in the \textit{proximal contact list}. 
 
 \item alert all the devices receiving the notification from the activated one, in order to warn about the possible contagion. The alert should contain precise directions on the actions to be taken (e.g. undergo a test and/or start a voluntary quarantine).
\end{enumerate}
The complete \textit{Proximity} workflow is depicted in Fig.(\ref{diagram}).\\

The crucial feature of \textit{Proximity} is the tracking of neighboring devices. This must occur offline and with high accuracy, for the measurement of the distance. The former requirement will ensure the users' \textit{privacy}, as all the information will be stored locally in the device, in an encrypted form. \par

Moreover, it will allow the app to operate even if the connection to the network is missing or weak. The range accuracy is crucial, in order to avoid having a large number of \textit{false alerts}. With false alerts we mean devices recorded in the \textit{proximal contact list}, which were out of the infection range (i.e. the maximum distance at which one can be infected by proximity). Clearly the closer the tracking range is to the infection range, the higher the identification accuracy of potentially infected people will be.

\par

We identified the \textbf{Bluetooth} technology as the most suitable option to satisfy the app's needs. Bluetooth allows for the direct communication of one device with multiple others, even out of the line-of-sight, thus enabling simultaneously tracking of multiple smartphones within the selected range. Furthermore, it is designed for low power consumption, enabling the app to be constantly active, without draining the battery. 
\par

Another crucial parameter is the working range, in order to provide the required accuracy. Current Bluetooth (Class 2) devices operating on mobile phones have a range of about 10 meters\cite{Bluetooth}, variable depending on the presence of obstacles along the path, which can attenuate the signal. The infection range is usually on the order of a few meters, so Bluetooth antennas are a viable tool for this application.\cite{Bluetooth}\par

One potentially interesting alternative to Bluetooth, would be \textbf{Near Field Communication} (NFC).
This is a pretty recent technology implemented in most novel devices, which allows for radio communication on very short distances, used for example for wireless payments. 

This is promising for \textit{Proximity} since it is a very low power consuming technology.\cite{NFC}
The drawback is the range, which currently is in the order of few
centimeters, thus making it unsuitable for our app, for the moment.

\subsection{Further development}
After listing above the main operating features of our \textit{Proximity} app, we propose here some further refinements. These improvements might help addressing some collateral issues linked to the use of the app, that we envision in an hypothetical epidemic scenario.\par

In the event of the activation of the \textit{Alert mode} for a verified infected person, the number of people who have been in contact with them, and thus receiving the alert and potentially willing to undergo the medical examination could be very high. The number of notifications sent will then increase with a cascade effect, damping the epidemic outbreak. Such scenario could put the health care system under pressure, due to the great number of tests requested almost simultaneously.
The solution we suggest is based on two different, but entangled, approaches.\par

The first relies on the fact that the app works as a detector of other devices.
In this context we can define strong and weak interactions. \par
The strength of the interaction is defined based on both the duration of the interaction itself and the distance between the interacting devices. The former can be measured by keeping track of the period of time two smartphones stay within the interacting range, while the latter can be estimated from the intensity of the Bluetooth signal at the receiving end. The duration of the interaction can also be used to track multiple contacts with the same device, by simply summing the times of the single interactions.

Introducing a level of intensity of the interaction permits to establish a priority scale, which is directly correlated to the probability of infection.
People who had longer interactions with an infected individual, will have a higher probability of incurring in the infection themselves, as well as people who have been much closer to an infected person. 

Once the priority is set, it is used to order the \textit{proximal contact list}. At this point, it is possible to proceed by defining a threshold, taking into account for instance the capacity of the health care system to perform tests for the detection of the infection. This way, if the number of people who can potentially receive the alert is higher than such threshold, then the individuals with lower priority are excluded from the list. They can, for example, be held in a waiting list, and alerted at a subsequent time. \\

The second approach that we suggest makes use of two different kind of alerts, operating in a hierarchical way. The first, named \textit{red notification}, corresponds to the notification system described above: once a person is tested positive, the \textit{proximal contact list} of their own device gets a notification of the contact. The second notification, \textit{yellow notification}, can be activated by a person who has received the \textit{red notification}, but not been tested yet. The \textit{yellow notification} is sent to the \textit{proximal contact list} of this second individual. In the latter case, the alert must contain indications that are different compared to those sent with the \textit{red notification}. For example, the app would only advise to take precautions but it would not request the medical examination, unless a further \textit{red notification} is received.\par

We recognize that this last solution might not always be possible or beneficial, because it could, for example, create more confusion in people receiving different alerts. We envision it as an optional functionality, whose activation can be authorized or not by the app provider, depending upon the specific conditions of the epidemic.

\section{Encryption and operating scheme}

In order to show how it is possible to guarantee privacy and anonymity, we describe in more detail the encryption system and the operating scheme we propose for our app.
\par
We believe that the optimal encryption scheme for an application like this is the \textit{asymmetric encryption}. \\
Asymmetric encryption involves a two-key system: a public key $ K_p $ which is shared between the sender and the receiver, and a private or secret key $ K_s $ which is known only to the receiver. \\
The sender, who knows the plain-text message $ n $, generates the encrypted message $ n ^ * $ through a "one-way function" $ Cr $ and a public key $ K_p $. \\
The receiver, who knows both $ K_p $ and $ K_s $ keys, is able to decrypt the message through $ Dr $, the decryption function.

\begin{equation}
\begin{array}{lllll}
  \qquad\quad K_p        \qquad\qquad\qquad\qquad\quad\quad\  K_p\\
  \qquad\quad \downarrow \qquad\qquad\qquad\qquad\quad\qquad   \downarrow \\
  n \longrightarrow \mathcircled{Cr}\longrightarrow n^* \quad \dashrightarrow \quad n^*\longrightarrow \mathcircled{Dr}\longrightarrow n\\
   \qquad\qquad\qquad\qquad\qquad\qquad\quad\quad\ \  \uparrow\\
   \qquad\qquad\qquad\qquad\qquad\qquad\quad\quad\ \  K_s\\
\end{array}
\end{equation}

The strength of this system is that, despite being $ K_p $ and $ K_s $ keys dependent on each other, the knowledge of $ K_p $ does not lead back to $ K_s $. Therefore, $ Cr $ and $ K_p $ alone do not allow the decryption of $ n ^ * $. \\
There are several algorithms that allow asymmetric encryption. The first of these, most famous and widely used, is the RSA algorithm \cite{RSA}.\\

\begin{figure*}[hbt!]
  \includegraphics[width=0.99\textwidth]{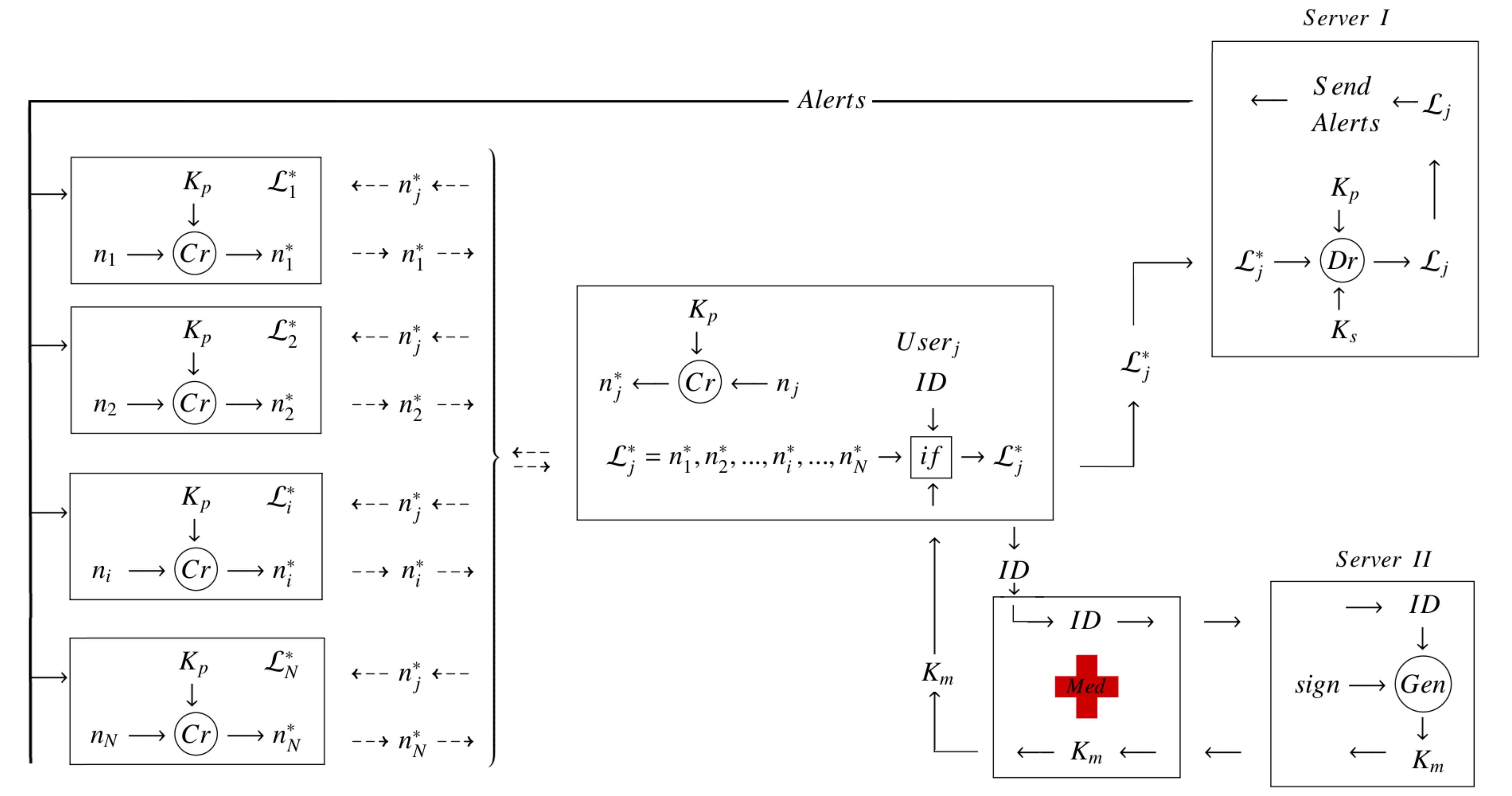}
    \caption{The diagram shows the operating scheme of the \textit {Proximity} app and the encryption system.}
    \label{cryptogram}
\end{figure*}

Asymmetric cryptography is the ideal ingredient to develop our recipe. Indeed, each user of the app can be considered as a sender of information (their phone number) in the encrypted form $n ^ * $. Users interacting with each other exchange this encrypted information, which is temporarily registered in the \textit{proximal contact list}:
\begin{equation}
\centering
\mathcal{L}^*_j=n^*_1,n^*_2, ...,n^*_i, ...,n^*_N  
\end{equation}
which is absolutely indecipherable for any user.

The whole set of users therefore represents an encrypted distributed database $ \mathcal {D} ^ * $ which turns out to be really challenging to break. \\
\begin{equation}
\centering
\mathcal{D}^*=\bigcup_i\mathcal{L}^*_i
\end{equation}

As previously specified, the user needs an additional key $ K_m $, which is provided by the medical authority, to enables the activation of the \textit{Alert mode}. Once the \textit{Alert mode} is active, the \textit{proximal contacts list} $ \mathcal{L}^* $ is sent to the decryption server $ Server \ I $, which eventually sends notifications to all the contacts present in the list.

The $ K_m $ key is generated at the moment of the request from the authority, by a different server $ Server \ II $, providing the user $ ID $ of the person tested positive and the name of the requesting doctor.
In this way, $ Server \ I $ only receives data from the $ j $ -th user, which is the one who is in possession of the $k_m$ key. This procedure adds an additional constraint that limits data sharing.\\

The intrinsic security of such a system relies first of all in the lack of a central database, since the data is distributed on all smartphones, in an encrypted form.
The only party able to decrypt the data is the one who generates the keys, who is also the owner of the system. However, it is important to underline that they do not own the data of all the users, but only the small part, of those who need to be alerted.
The asymmetric encryption algorithm can only be violated with an incredible computational cost.\cite{RSA} However, further improvements, like increasing the size of the keys to a number greater than $2048 $ bits and changing the public key often (e.g. daily), would make the system safe from any possible reasonable attack.

\section{Monte Carlo simulation}

Intuitively, it is clear that promptly warning all the proximal contacts of an infected person allows for slowing down the epidemic.
Nonetheless, in order to show the app efficiency we modeled the functioning of the app using a simplified Monte Carlo simulation. We have
not modeled the entire evolution of the epidemic, since we are only interested in the initial phase, the exponential growth.\par
Inspired by the SIR model \cite{SIR}, we introduce the quantity $R_0$ the "basic reproductive rate" as the expected number of secondary cases directly generated by a single positive case during the incubation time $\Delta T_{inc}$.
The quantity $R_0$ represents the crucial parameter which describes the epidemic evolution.\\

The idea is that by acting on this coefficient $R_0$, it is possible to slow down and even stop the epidemic. This is exactly what the app aims to do; by promptly alerting the infected people, with an \textit{adaptive confinement} conceptually described in Fig.(\ref{confinement}), the rate of epidemic growth is significantly reduced.

The assumptions for the Monte Carlo numerical simulation is the following:
\begin{itemize}

\item the average incubation period of the disease $ \Delta T_{inc}$ is constant.

\item during the incubation period, the "basic reproductive rate" $R_{0}$ is constant.

\item the incubation period $\Delta T_{inc}$ is equivalent to the duration of infectiousness. We assume that once the symptoms appeared the patient is strictly quarantined. Therefore, for any time $t>\Delta T_{inc}$ we set $R_{0}^{\ Positive}=0$.

\item once an individual receives the alert from the app, they can start a voluntary quarantine, which decreases $R_0$ of a factor $k$, $R_{0}^{\  Alert}=R_0/k$, and undergoes the test. 

\item once the epidemic is ongoing, we assume that the activation of the app
into the \textit{Alert mode} from the $i$-th day (corresponding to $T_{ith}$). Until the $i$-th day, the app is in a silent mode: just tracking but not alerting.

\item  we assume a linear gradual diffusion of the alert during the days following the $i$-th day (this period is defined $T_{delay}$), in order to roughly model two events: the progressive mitigation of the sudden test request, which can hardly be satisfied by the health care system, and the gradual increase of the app users number.
\end{itemize}

\begin{figure}[htbp!]
\centering
\includegraphics[width=0.49\textwidth]{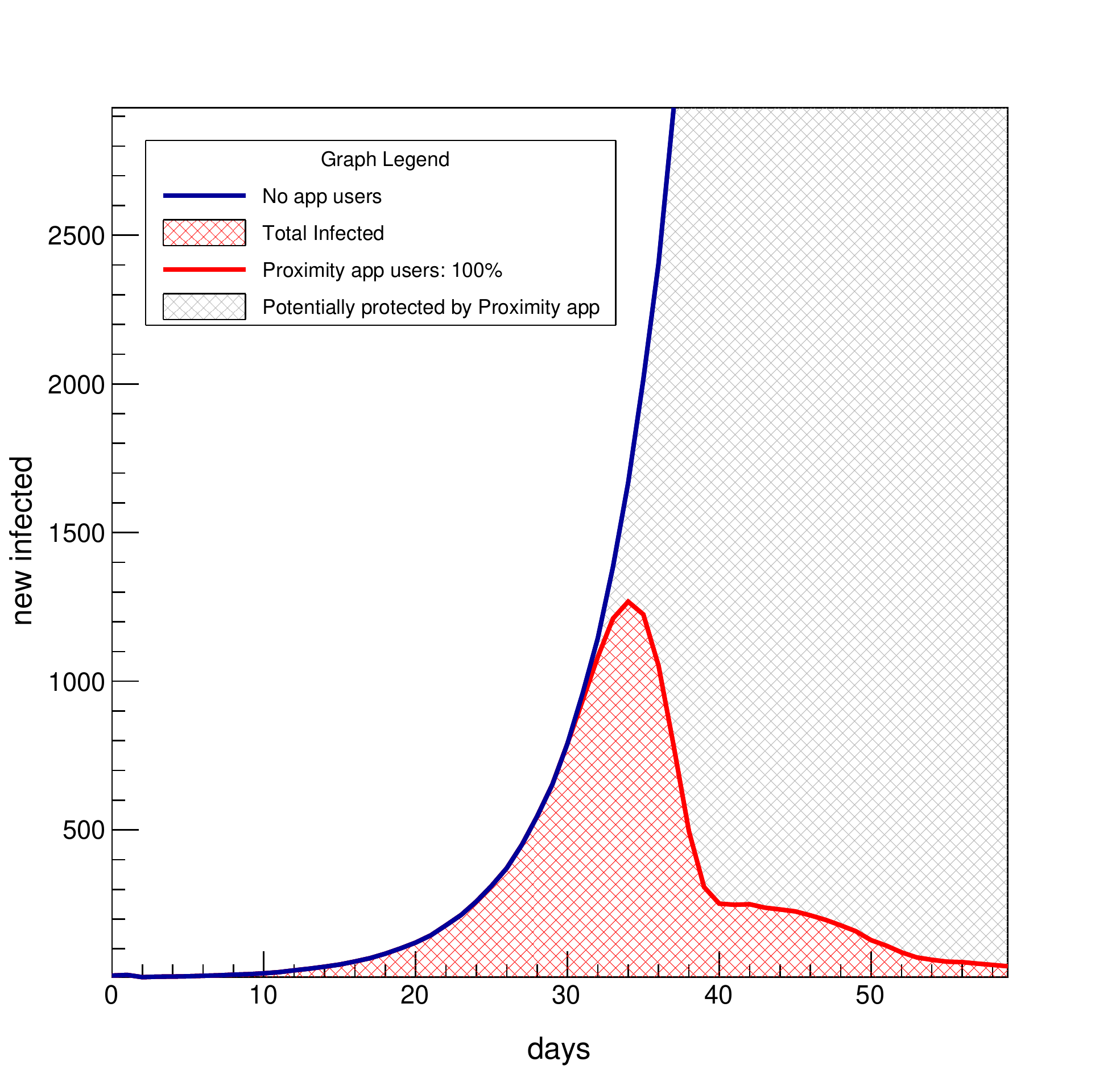}
\caption{Daily new infected trend according to the model described in the text. The blue line is the epidemic growth and the red line is the simulated evolution with the Proximity app. The activation starts from the $30th$ day. The grey area represents the number of people potentially protected. The parameters of the simulation are: $\Delta T_{inc}=14$ days, $R_{0}=3$, $k=10$, $T_{ith}=30$, $\Delta T_{delay}=10$ days. Data have been averaged over 50 simulations.}
\label{graphic1}

\end{figure}

We believe that these assumptions can sufficiently describe the dynamics of the spread of the infection under the effect of the app.\\
The results are shown in Fig.(\ref{graphic1}), the grow of the infection is exponential as expected, and from the $i$-th day the app starts to be activated gradually and the outbreak drops dramatically.\\
According to these simulations, the app's intrinsic efficiency, which is based on the adaptive 

confinement/quarantine, seems to be excellent.\par

On the other hand, the silent assumption that all the infected individuals would be users of the app is not realistic, as well as the assumption that all the possible infected people could be on the proximity list.\par

\begin{figure}[htbp!]
\centering
\includegraphics[width=0.49\textwidth]{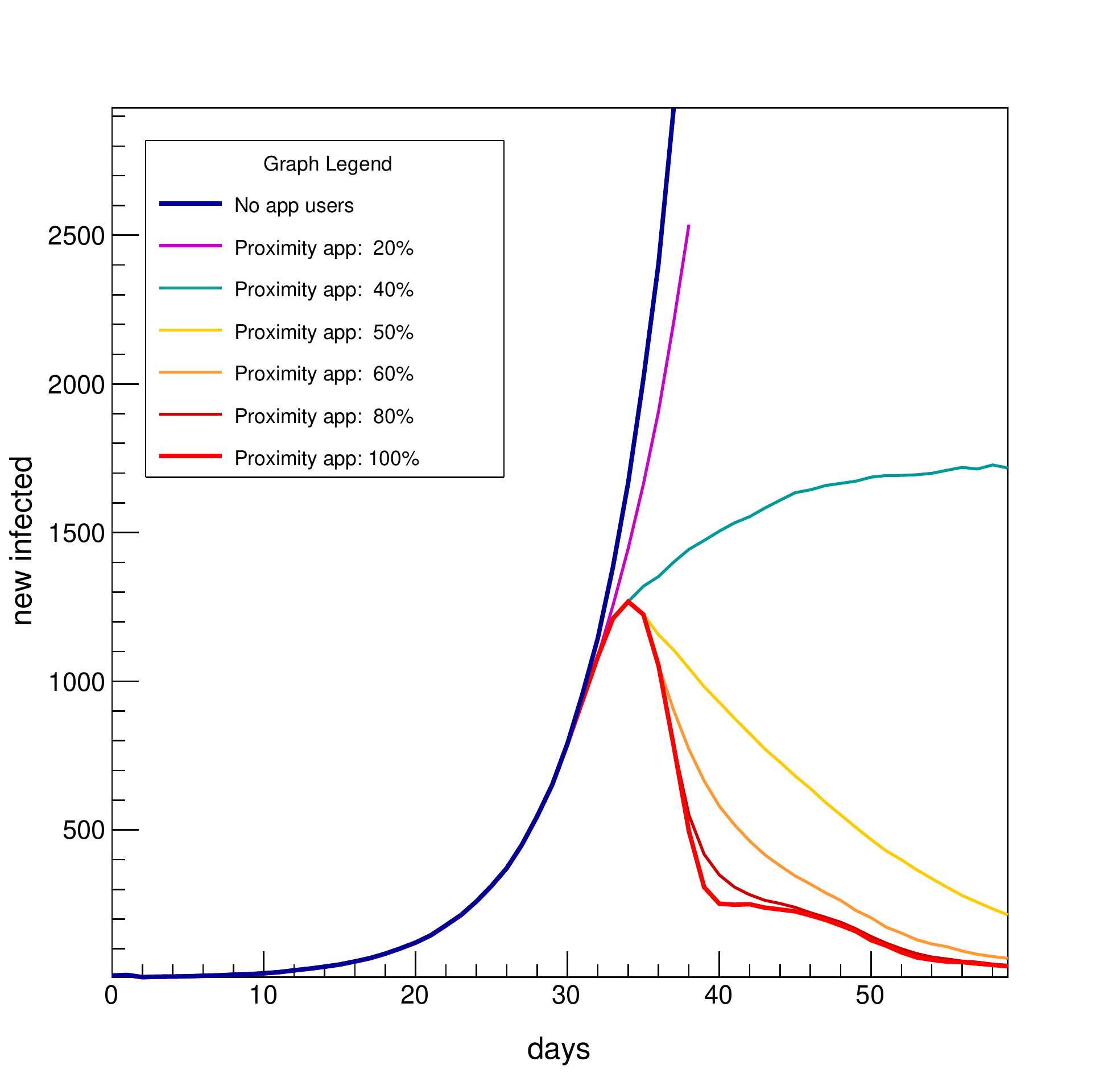}
\caption{Daily new infected trend according to the model described in the text. Several scenario were evaluated with different values of efficiency. It is relevant to highlight that with an efficiency of about $60\%$ the app could slow down the outbreak very well.
The parameters of the simulation are: $\Delta T_{inc}=14$ days, $R_{0}=3$, $k=10$, $T_{ith}=30$, $\Delta T_{delay}=10$ days.
Data have been averaged over 50 simulations.}
\vspace{3mm}
\label{graphic2}
\end{figure}

Therefore, in Fig.(\ref{graphic2}) we show the results of several simulations with a lower efficiency. \par
The efficiency could be interpreted in at least two ways, or a combination of both: 
\begin{itemize}
\item the fraction of app users with respect to the totality of infected,
\item the fraction of infected individuals that can not be reached/tracked by the app due to technical reasons, for instance: smartphones shutdown or indirect contacts such as those through contaminated surfaces.\par
\end{itemize}
Independent of the former interpretations adopted for the efficiency, it is relevant to highlight that, according to our model, with an efficiency of about $60\%$ the app
could still slow down the outbreak very well.\\

The $ k $ parameter represents the effectiveness of the quarantine of those  who know they have been exposed to a

risk of infection, those who have received the alert.\\

\begin{figure}[htbp!]
\centering
\includegraphics[width=0.49\textwidth]{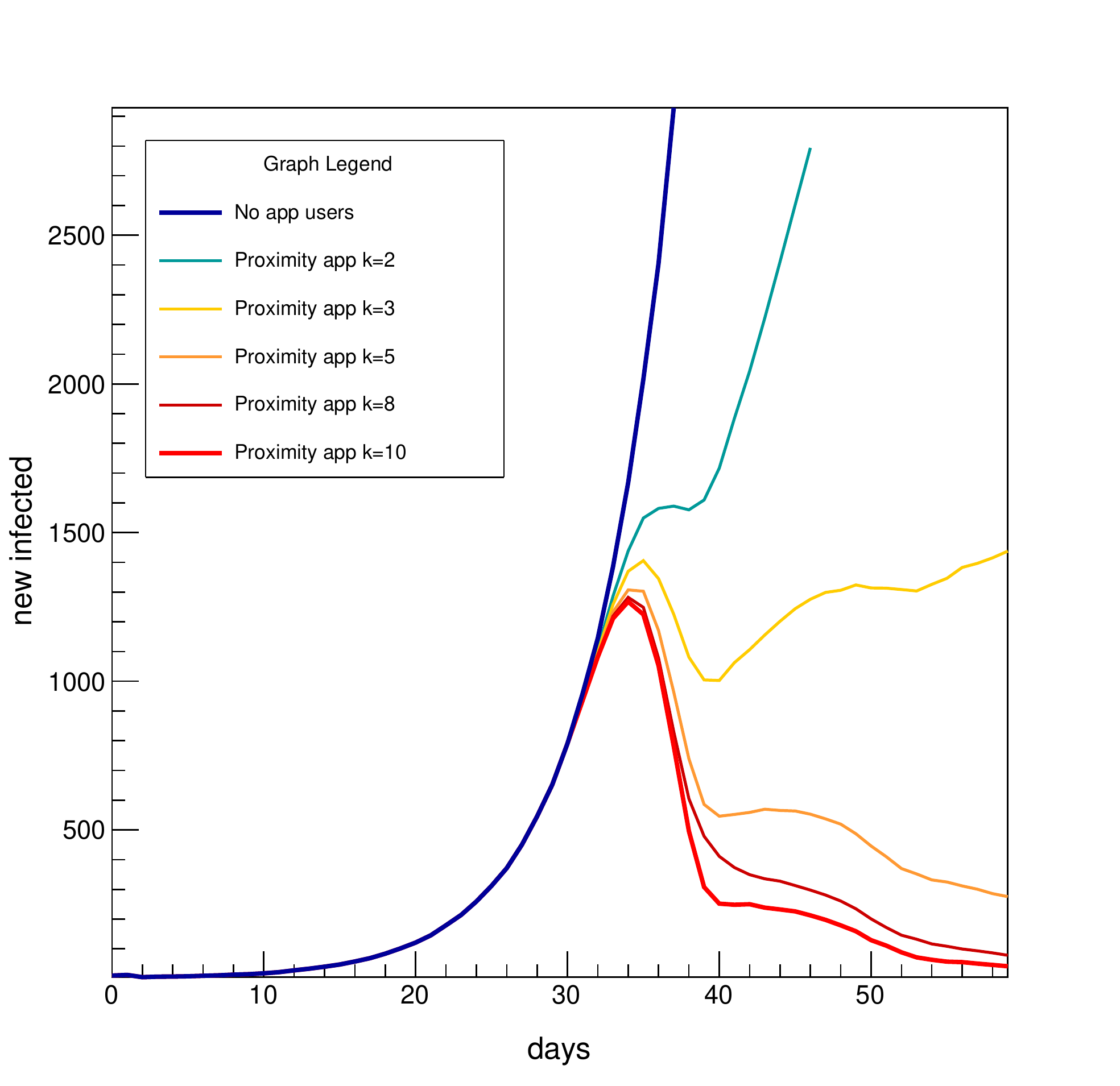}
\caption{Daily new infected trend according to the model described in the text. Several scenario were evaluated with different values of $k$. It is possible to interpret this parameter as the effectiveness of the quarantine of those who know they have been exposed to a contagion risk. Assuming sensible and prudent behavior we expect $ k \gg 1 $.
The parameters of the simulation are: $\Delta T_{inc}=14$ days, $R_{0}=3$, $T_{ith}=30$, $\Delta T_{delay}=10$ days.
Data have been averaged over 50 simulations.}
\label{graphic3}
\vspace{1mm}
\end{figure}

If the alert is completely ignored, the value of the parameter would be $ k = 1 $, instead we expect many people to consider seriously the contagion risk notification and intensify the voluntary quarantine and undergo the test as soon as possible.\\
These precautions involve an increase in the parameter $ k $ which we assume is probably around the value $ k = 10 $.
In Fig. (\ref{graphic3}) we studied the dependence of our model on changing the parameter $ k $.\\

The $ T_ {delay}$ parameter represents the effectiveness and quickness of the national health system (NHS) in testing and communicating the outcome to positive patients. The delay is considered from the day when the Alert activation protocol is introduced to full operation.\\
Another cause of this possible delay is the gradual increase in the number of users of the app.
We probably expect a week, ten days. \par
\begin{figure}[htbp!]
\centering
\includegraphics[width=0.49\textwidth]{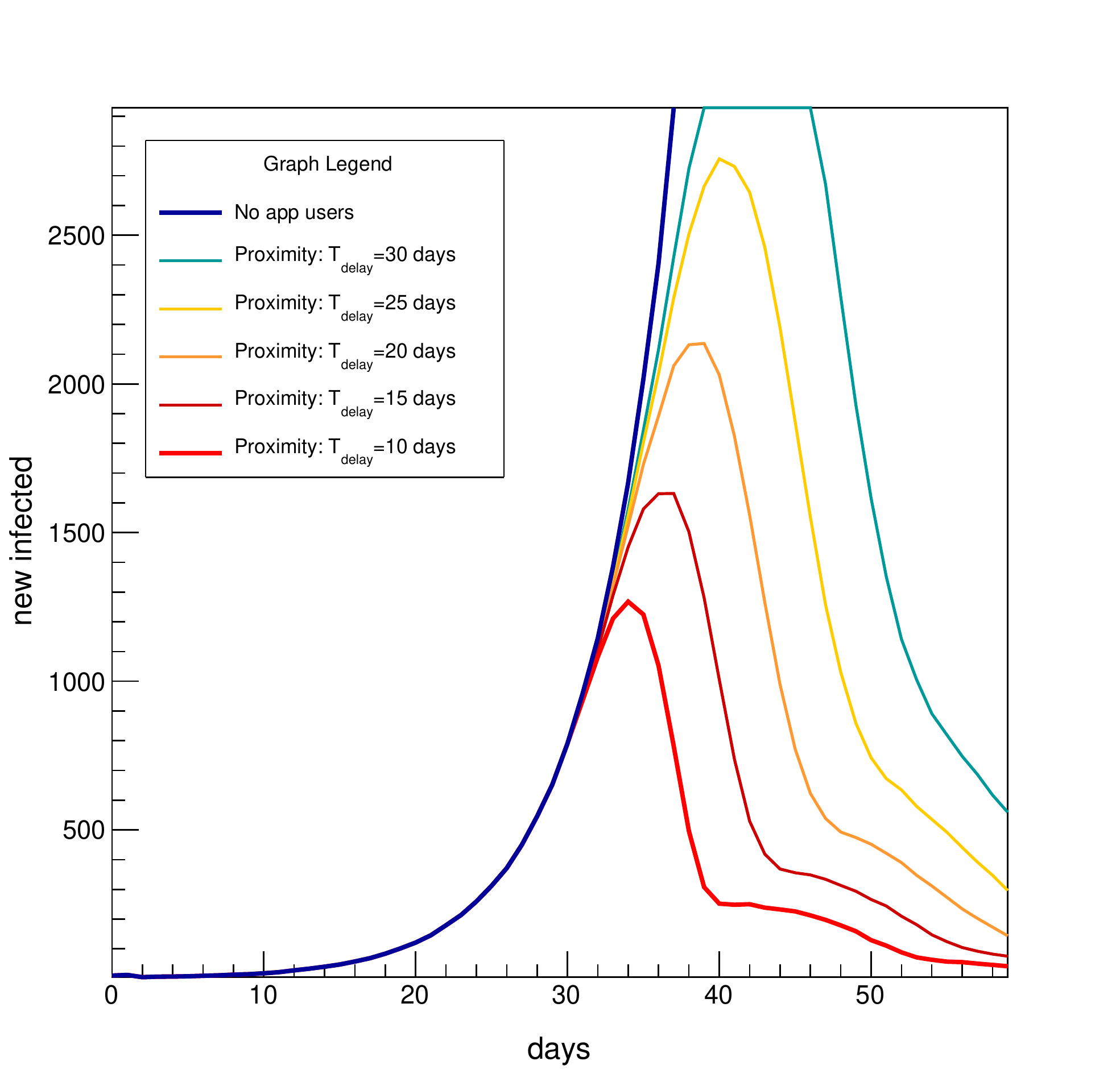}
\caption{Daily new infected trend according to the model described in the text. Several scenario were evaluated with different values of $ \Delta T_ {delay}$.
It is possible to interpret this parameter as the effectiveness of the NHS in testing patients and promptly reporting the positives.
The parameters of the simulation are: $\Delta T_{inc}=14$ days, $R_{0}=3$, $k=10$, $T_{ith}=30$.
Data have been averaged over 50 simulations.}
\label{graphic4}
\end{figure}

In Fig. (\ref{graphic4}) we studied the dependence of our model on changing the parameter $ k $ \par

\section{Discussion}
Although to contain the epidemic other apps have been proposed \cite{stit} or are under construction, we believe that our \textit{Proximity} app recipe gathers a large number of advantages.
In this section we will try to highlight its strengths and its weaknesses that require a further development.\\
\begin{enumerate}
\item All the information about the interactions is temporarily collected by the app in an \textbf{encrypted list}. \\

\item The app does not require any central database, all the data are distributed in the \textbf{local storage} of the smartphones. Only the \textit{proximal contact list}
of people certified as infected are sent to a remote server to be decrypted to allow contacts to be notified anonymously.\\

\item Proximal contacts are detected directly and with an \textbf{offline system}: there is no need of an internet connection at this stage.
\end{enumerate}

\newpage

\begin{enumerate}
\setcounter{enumi}{3}

\item A direct detection of the proximal contact is potentially \textbf{more accurate than the geolocation}, and it can be carried out in places where satellite coverage is absent or scarce. Thanks to the intensity of the signal, the proximity radius can be tuned more precisely.\\ 
Further development of mobile phones technology can help to improving these aspects.\\
\item The possibility of alerting unknown people directly and in \textbf{complete anonymity}. \\
\item To avoid spreading unwanted or fake alerts, the activation of the alert can happen only thanks to a key provided by the \textbf{medical authority} who certifies the case of epidemic contagion.\\
\item The app is relatively simple, \textbf{low energy consuming}, low memory and low computational cost are required. On the other hand, it requires at least a solid encryption system, the development of a brand new communication system with the current technology available (Bluetooth) and an anonymous way to send the alert (preferably via a SMS).\\
\item The app provides a system of \textbf{priorities} based on the duration and the distance of the contact, which enhances its efficiency and lowers the numbers of false alerts.\\
\item The app can be disabled when there is no ongoing epidemic.
A warning to switch on the app should be send in case of epidemic risk. Furthermore, the app can be temporarily disabled by the user or paused at his discretion.\\
\end{enumerate}

\section{Conclusions}

We believe that protecting humanity from epidemics is of crucial importance and of interest to all.\\
For this reason, we are sure that a system like the one we have proposed will inevitably be adopted in the future.
The hope is that this will happen in the shortest time possible and especially in respect of rights like privacy and freedom. 
We believe this solution may be useful in particular in the current COVID-19 pandemic and moreover could be used to prevent similar events in the future.

\section{Statement}
The authors claim to have developed the content of this paper independently without taking inspiration from others.\\
Despite this, they recognize in the \textit{TraceTogheter} app \cite{TT} presented in Singapore on March 20, 2020 a realization very similar to what is described in this work, albeit with some important differences.
Proximity\cite{Prox} does not require a centralized database, whereas this is envisaged in \textit{TraceTogheter} \cite{TT}.

\section{Acknowledgements}
We thank Colin Pavan and Joel Klinger for useful suggestions and comments and for reviewing the English version and Clémence Doliget for the French translation.


\begin{thebibliography}{99}
\bibitem{BillGates} Bill Gates, "The next outbreak? We’re not ready", TED Talks 2015
\bibitem{SIR}Hethcote H (2000). "The Mathematics of Infectious Diseases". SIAM Review. 42 (4): 599–653
\bibitem{WHO} "Q\&A on coronaviruses". World Health Organization (WHO). 11 February 2020.
\bibitem{paz0} Mohammadi, D. (2019). Finding patient zero. Acute pain, 10, 00.
\bibitem{Bluetooth} "Basics $|$ Bluetooth Technology Website". Bluetooth.com. 23 May 2010.
\bibitem{NFC}"Near Field Communication Versus Bluetooth". 28 November 2012.

\bibitem{RSA} Original RSA Patent. Rivest; Ronald L. (Belmont, MA), Shamir; Adi (Cambridge, MA), Adleman; Leonard M. (Arlington, MA), December 14, 1977, U.S. Patent 4,405,829

\bibitem{Prox}"Proximity: a recipe to break the outbreak". M. Faggian, M. Urbani, L. Zanotto. 23 March 2020. \\
https://arxiv.org/abs/2003.10222

\bibitem{Epi} Yoneki, Eiko, and Jon Crowcroft. "Epimap: Towards quantifying contact networks for understanding epidemiology in developing countries." Ad Hoc Networks 13 (2014): 83-93.

\bibitem{stit} Stopcovid19 app : https://www.stopcovid19.it/it/

\bibitem{TT} TraceTogheter app: www.tracetogether.gov.sg

\end{thebibliography}
\end{document}